\documentclass[%
 twocolumn,
 amsmath,amssymb,
 aps,
pra,
]{revtex4-1}

\usepackage{graphicx}
\usepackage{dcolumn}
\usepackage{bm}
\usepackage{epsfig}
\usepackage{epstopdf}
\usepackage{xcolor}

\begin{document}


\title{Enhancing the capture velocity of a Dy magneto-optical trap with two-stage slowing}

\author{William Lunden, Li Du, Michael Cantara, Pierre Barral,  Alan O. Jamison, and Wolfgang Ketterle}
\affiliation{Research Laboratory of Electronics, MIT-Harvard Center for Ultracold Atoms, Department of Physics,
Massachusetts Institute of Technology, Cambridge, Massachusetts 02139, USA}%

\date{\today}

\begin{abstract}
Magneto-optical traps (MOTs) based on the $626\;{\rm nm}$, $136\;{\rm kHz}$-wide intercombination line of Dy, which has an attractively low Doppler temperature of $3.3\;\mu{\rm K}$, have been implemented in a growing number of experiments over the last several years. A challenge in loading these MOTs comes from their low capture velocities. Slowed atomic beams can spread out significantly during free-flight from the Zeeman slower to the MOT position, reducing the fraction of the beam captured by the MOT. Here we apply, for the first time in a Dy experiment, a scheme for enhancing the loading rate of the MOT wherein atoms are Zeeman-slowed to a final velocity larger than the MOT's capture velocity, and then undergo a final stage of slowing by a pair of near-detuned beams addressing the $421\;{\rm nm}$ transition directly in front of the MOT. By reducing the free-flight time of the Zeeman-slowed atomic beam, we greatly enhance the slowed flux delivered to the MOT, leading to more than an order of magnitude enhancement in the final MOT population.
\end{abstract} 

\maketitle


\section{Introduction}
Dysprosium, which possesses the largest magnetic moment ($\mu\approx 10 \mu_B$) of any atomic species, has grown in popularity in the ultracold quantum gas community over the last decade \cite{TwoStageMOT,PfauMOT,ItalyMOT,ParisMOT,MainzMOT,ParisMOT,FerlainoMOT,Supersolid1,Supersolid2,Droplet1,Droplet2,LevSOC} The large magnetic moment, as well as several other useful properties, arise from its [Xe]$4f^{10}6s^2$ electronic configuration. The two $6s$ electrons give rise to a helium-like excitation spectrum, including a strong transition at $421\;{\rm nm}$ and a weak, intercombination transition at $626\;{\rm nm}$. The unfilled $4f$ shell gives rise to narrow clock-like transitions. It also leads to spin-orbit coupling in the ground state, which is useful for many quantum simulations, including simulating gauge fields  \cite{LevGauge,LevSOC}.

The narrow linewidth of the $626\;{\rm nm}$ transition in Dy corresponds to a low Doppler temperature of $3.3\; \mu {\rm K}$, making it an attractive option for magneto-optical trapping. The downside of using a narrow transition is that the capture velocity of the MOT is lower than for a broader transition. Slowing an atomic beam to within a low capture velocity can lead to a situation where the slowed beam transversely spreads out so much that many slowed atoms miss the MOT. Cooling the transverse degrees of freedom of the atomic beam and increasing the capture velocity of the MOT by frequency-dithering the MOT light are two measures which are typically employed to mitigate this limitation \cite{MainzMOT,ItalyMOT,PfauMOT,ParisMOT,LevMOT2,FerlainoMOT}, but their effectiveness can be limited.

In the present work, we add a new approach which we refer to as ``angled slowing,'' which applies a second stage of slowing to the atomic beam with a pair of low-power beams that intersect directly in front of the MOT. This allows us to choose a sufficiently large final velocity for the first stage of slowing (i.e., Zeeman slowing) that atoms do not spread out appreciably before reaching the MOT. This approach was introduced in an Yb experiment, where it gave a small enhancement to the MOT loading rate \cite{BenThesis}. In our experiment, angled slowing enhances the MOT population by more than a factor of 20. Compared to other methods which have been employed to increase the capture velocity of narrow-line MOTs --- such as the two-stage MOT \cite{TwoStageMOT} and the core-shell MOT \cite{CoreShell}--- the angled slowing approach requires fewer beams and less laser power.

In section II we briefly describe the aspects of our experiment that are similar to those previously reported by other experiments. In section III, the idea behind angled slowing and how it is particularly applicable to experiments with narrow-line MOTs is discussed. In section IV, we describe how we optimized the performance of angled slowing with respect to beam pointing, laser power, and frequency. In section V, the compression and detection sequence that follows the loading of our MOT is described, and the temperature and phase space density of the compressed MOT are reported.

\section{Experimental Setup}

\begin{figure*}
    \includegraphics[scale=0.2]{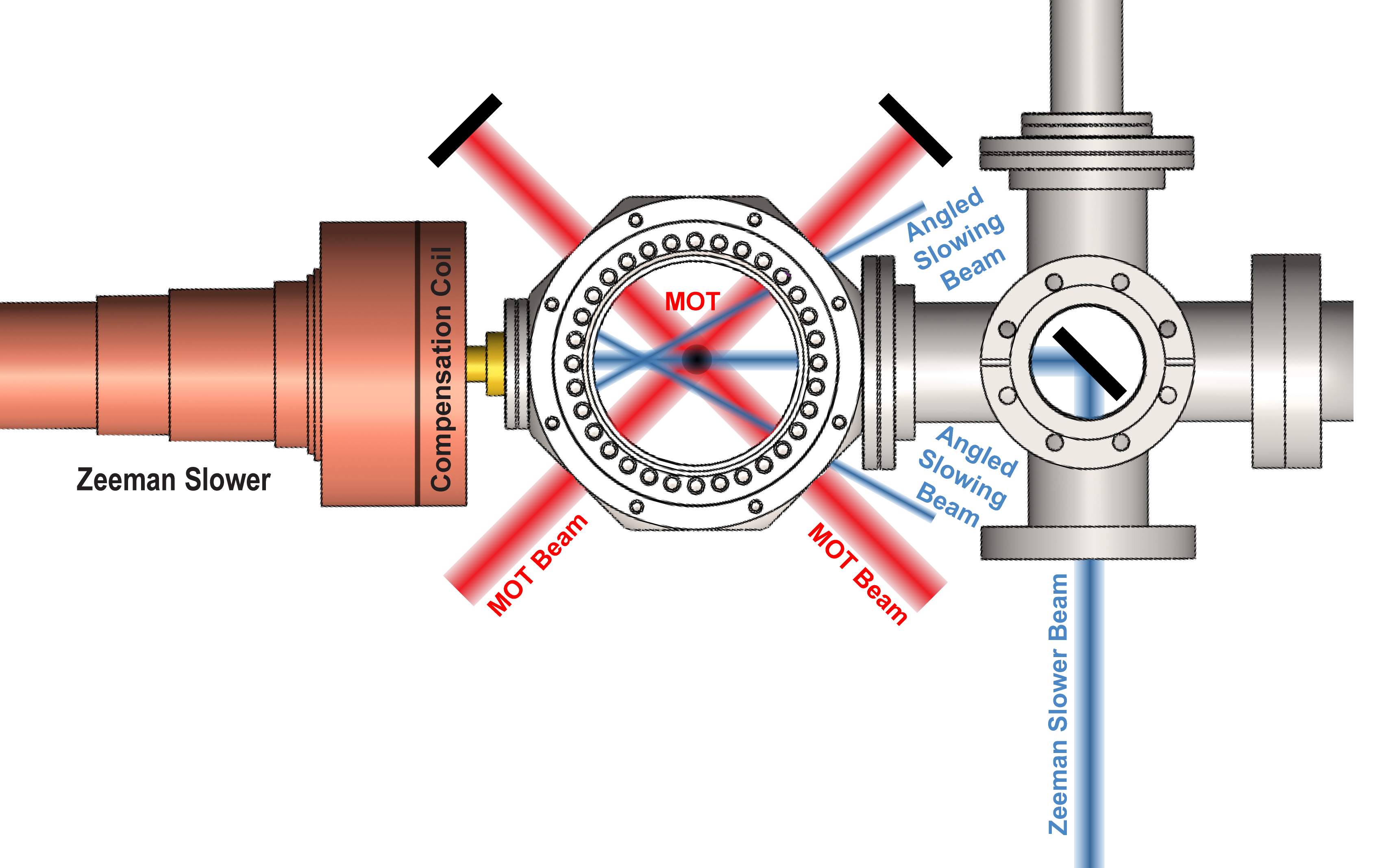}
    \caption{Geometry of our Zeeman slowing, MOT, and angled slowing beams. The vertical MOT beam pair is not shown in this top view of the machine. The angled slowing beams enter through the same viewports as the MOT beams, and are aligned such that they intersect the atomic beam without hitting the MOT (indicated by the dark circle in the center of the chamber).}
    \label{fig:TheMachine}
\end{figure*}

Due to the recent explosion in popularity of dysprosium, several groups have developed similar cooling and trapping protocols in parallel \cite{MainzMOT,ItalyMOT,PfauMOT,ParisMOT,LevMOT2,FerlainoMOT}. Here we briefly describe our approach, and give references to more detailed explanations of similar systems.

Our atomic beam of Dy is generated by a commercial molecular beam epitaxy oven \footnote{SVT Associates model SVTA-DF-20-450} heated to $1250^\circ {\rm C}$. The dysprosium vapor is collimated into an atomic beam by a $7\;{\rm mm}$ diameter nozzle, which is $90 \;{\rm mm}$ from the opening of the oven, followed by a $10\;{\rm cm}$ long, $7\;{\rm mm}$ diameter differential pumping tube that starts $19\;{\rm cm}$ from the nozzle.

We use $421\;{\rm nm}$ laser light for Zeeman slowing, transverse cooling, and absorption imaging. This light is generated using an M-Squared Ti:Sapphire laser and ECD-X fixed-frequency second-harmonic generation cavity ($1.6\;{\rm W}$ total output), as well as two injection-locked laser diodes ($90\;{\rm mW}$ total output each). The frequency of the Ti:Sapphire laser is stabilized by measuring the frequency of the doubled light with a HighFinesse WLM-7 wavemeter, and feeding back on a piezo-actuated mirror in the laser cavity using an Arduino Due. We periodically adjust for drifts in the calibration of the wavemeter (which are typically a few MHz per day) by checking the resonance frequency of the $421\;{\rm nm}$ transition via absorption imaging. A Toptica TA-SHG system ($700\;{\rm mW}$ total output) generates the $626\;{\rm nm}$ light for the MOT. For frequency stabilization of this laser, we shift the light by about $+1\;{\rm GHz}$ and employ a modulation transfer spectroscopy scheme to lock the laser frequency to a transition in a room-temperature iodine cell.

In the present work we slow the bosonic isotope $^{162}$Dy, which has 25.5\% natural abundance \cite{MaierThesis}. Our Zeeman slowing light consists of 300mW of light addressing the $421\;{\rm nm}$ transition ($ \Gamma_{421} = 32.2\; {\rm MHz}$), which comes to a focus at the position of the oven. Light enters the vacuum chamber with a beam diameter of about $2\; {\rm cm}$, bouncing off of a 45-degree in-vacuum mirror as shown in Figure 1. This scheme was implemented so the entrance window for the slowing light does not get coated by the Dy atomic beam.

To minimize the effect of the Zeeman slower light on atoms trapped in our MOT, we use an increasing-field Zeeman slower design. This allows us to employ a larger detuning in our slowing beam, which reduces the losses due to scattering in the MOT. A counter-wound segment of coils at the end of the slower cancels the fringing magnetic field from the slowing coils at the position of the MOT. We use light detuned about $1.1\;{\rm GHz}$ from the zero-velocity transition, which resonantly addresses atoms moving at $480\;{\rm m/s}$ (close to the most probable velocity of the atoms emitted from the oven). We have an additional, uniformly wound bias coil running the length of the Zeeman slower, which creates a constant offset magnetic field inside the slowing region. This allows adjustment of the effective detuning of the Zeeman slower beam by up to several hundred MHz without needing to employ an accousto-optic modulator (AOM). 

Our MOT is formed by three retroreflected $626\;{\rm nm}$ beams. Each beam has a diameter of $2.3\;{\rm cm}$ and a total power of $42\; {\rm mW}$ ($\pm 5\%$), corresponding to a (peak) saturation parameter of $s \approx 280$. The quadrupole field's gradient along the strong direction is approximately $2.5\;{\rm G/cm}$. To improve the capture velocity of the MOT we dither the frequency of the MOT light using a double-passed AOM. The dithering occurs at a frequency of $120\; {\rm kHz}$ and broadens the laser linewidth to $2.6\;{\rm MHz}$ ($30 \Gamma_{626}$). Three pairs of rectangular coils in Helmholtz configuration allow us to cancel background magnetic fields, and will also allow us to employ feedback- and feedforward-based magnetic field stabilization schemes during future experiments.

\section{Angled Slowing}

Atoms that have been slowed by a Zeeman slower must travel some nonzero distance at their final, slowed velocity from the end of the Zeeman slower to the position of the MOT. During this period of free flight, the transverse velocity distribution of the beam causes the atomic beam to spread out. If the free flight time is sufficiently long, then the atoms can spread out far enough that they are not captured by the MOT.

While this is not typically a limiting factor in experiments, the combination of an increasing-field slower and the narrow linewidth of the MOT transition creates a situation in which the transverse spread plays a significant role. To clarify this point, we compare Dy to the more common alkali MOTs.

The capture velocity of a MOT can be estimated by calculating the largest velocity that can possibly be slowed to a stop within the profile of the MOT beams. Assuming that the atoms scatter photons at the maximum possible rate $\frac{\Gamma}{2}$ across an entire beam diameter $D$, an expression for the capture velocity is given by
\begin{equation}
    v_\text{cap} = \sqrt{2 \frac{\hbar k \Gamma}{2 m} D}
\end{equation}
where $m$ is the atomic mass and $k = \frac{2 \pi}{\lambda}$ is the wavenumber of the MOT light.

The spatial spread $\sigma$ of the atomic beam can be estimated as
\begin{equation}
    \sigma \approx 2 d \frac{v_\text{trans}}{v_\text{long}}
\end{equation}
where $d$ is the free-flight distance, $v_\text{trans}$ is the RMS transverse speed, and $v_\text{long}$ is the average longitudinal speed. Collimation of the atomic beam by one or more apertures typically leads to a transverse velocity distribution with an RMS speed around 1\% of the average (unslowed) longitudinal velocity \footnote{When transverse Doppler cooling of the atomic beam is employed prior to Zeeman slowing, the transverse velocity distribution can become much narrower; however, the light scattering during Zeeman slowing causes the transverse velocity distribution to re-broaden.}. 

Let us compare the case of a $^{87}$Rb MOT to a $^{162}$Dy MOT, taking typical values of $D = 2\;{\rm cm}$ for the MOT beam diameters. For Rb, $\lambda = 780\;{\rm nm}$, $m = 87\;{\rm amu}$, and $\Gamma = 2\pi \times 6\;{\rm MHz}$. This corresponds to a capture velocity of $67\;{\rm m/s}$. A typical initial most-probable velocity for atoms effusing from a Rb oven is about $330\;{\rm m/s}$, and so $3.3\; {\rm m/s}$ is a reasonable estimate of the RMS transverse speed of the atoms. If we consider an atomic beam slowed to the capture velocity value and an example free-flight distance of $10\; {\rm cm}$ we can estimate the spread of the atomic beam to be
\begin{equation}
    \sigma_\text{Rb} \approx 1\;\text{cm}
\end{equation}
which is smaller than the size of the MOT beams.

For a Dy MOT, $\lambda = 626\;{\rm }$, $m  162\;{\rm amu}$, and $\Gamma_{626} = 2\pi \times 136\;{\rm kHz}$. This gives a capture velocity of $8\;{\rm m/s}$. The most probable velocity of the Dy atoms effusing from our oven is about $480\;{\rm m/s}$, so $4.8\;{\rm m/s}$ is a reasonable estimate of the average transverse speed. For a free-flight distance of $10\;{\rm cm}$, we estimate the spread of the atomic beam to be
\begin{equation}
    \sigma_\text{Dy} \approx 12\;\text{cm}
\end{equation}
which is much larger than the size of the MOT beams. We thus see that the narrow linewidth of the $626\;{\rm nm}$ transition already lead to a significantly larger spreading of the atomic beam than in a typical alkali MOT.

Employing an increasing-field slower, while effective in reducing scattering losses in the MOT due to the larger Zeeman slower laser detuning, further exacerbates the transverse spreading problem. One reason is that the larger detuning reduces the amount of off-resonant slowing that occurs during the free-flight distance. The more off-resonant slowing that occurs during the free-flight, the larger the initial exit velocity from the Zeeman slowing region can be. We can estimate the typical effect of off-resonant slowing in a Dy experiment: a typical Zeeman slower beam detuning in a spin-flip slower is around $-18\Gamma_{421}$, with (resonant) saturation parameters of $s_0\approx 1$ \cite{PfauMOT}. If we assume that the slowed atoms scatter at a (detuned) saturation parameter of $s = \frac{s_0}{1 + 4\Delta^2/\Gamma^2} \approx 7.7\times 10^{-4}$ over a $10\;{\rm cm}$ free-flight distance, then we can estimate that atoms with exit velocities as high as $13 \;{\rm m/s}$ will be decelerated to within the capture velocity of the MOT.

A second reason is the increased free-flight distance due to the need for field-cancelling coils near the MOT. In an increasing-field Zeeman slower, the largest numbers of windings are closest to the MOT. As a result, it is necessary to compensate for the large fringing fields with an oppositely-wound compensation coil, so that the total residual magnetic field and field curvature at the position of the MOT is close to zero. Slowed atoms must thus travel an extra distance of several cm compared to the travel distance in spin-flip Zeeman slowers. In our experiment, the free-flight distance is $16\;{\rm cm}$.

\begin{figure}
    \centering
    \includegraphics[scale=0.35]{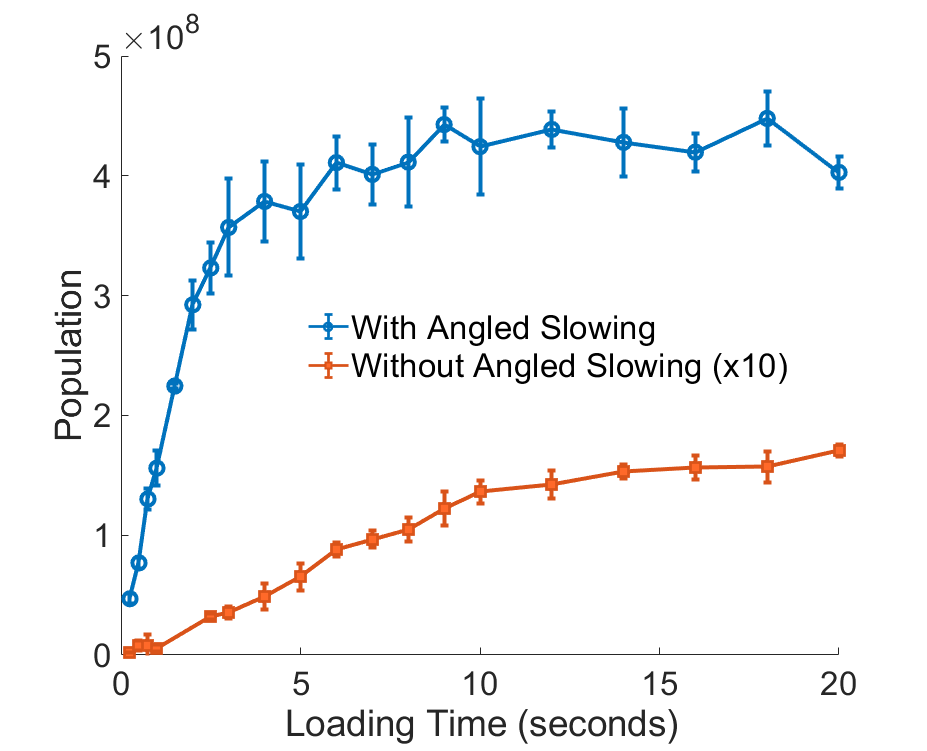}
    \caption{Population of the MOT as a function of loading time. The blue curve (open circles) shows the population when both transverse cooling and angled slowing with the optimal parameters of $\delta = -50\;{\rm MHz}$ and $7\;{\rm mW}$ per beam are employed. The orange curve (filled squares) shows the population when both the transverse cooling and angled slowing beams were turned off, multiplied by a factor of ten for visual clarity.}
    \label{fig:MOTloading}
\end{figure}

The purpose of the angled slowing scheme is to reduce the free-flight time by allowing atoms to exit the Zeeman slower at velocities well above the MOT's capture velocity. A few cm before the MOT, two beams with red detuning on the order of $\Gamma$ intersect the atomic beam to provide a net longitudinal slowing force, slowing the atomic beam to within the capture velocity of the MOT. The transverse components of the two beams' scattering forces are oppositely oriented and thus cancel. In effect, the addition of the angled slowing beams increases the capture velocity of the MOT. 

The advantage of using a pair of angled beams over a single beam colinear with the main Zeeman slower light, or adding a near-resonant sideband to the Zeeman slower, is that scattering losses in the MOT are avoided. Angled slowing also requires fewer beams and less laser power than the recently reported core-shell MOTs for alkaline-earth-like atoms \cite{CoreShell}. The setup for the angled slowing beams is depicted in Figure 1.

Without employing angled slowing, optimization of our Zeeman slowing parameters led to a MOT population of about $10^7$ atoms. We observe more than a factor of 20 gain in the final population of our MOT when using angled slowing. As described in the next section, we found optimal angled slowing performance with only $7\;{\rm mW}$ per beam, and a detuning of $-50\;{\rm MHz}$ ($-1.6 \Gamma_{421}$). The beam diameters are about $5\; {\rm mm}$, putting us far below the saturation regime ($I_{sat} = 56 \; \text{mW}/\text{cm}^2$). Figure 2 shows the population with and without angled slowing as a function of MOT loading time.

\section{Optimization of Angled Slowing}

We determined the optimal alignment of our angled slowing beams by maximizing the steady-state MOT population as measured by the integrated $626\;{\rm nm}$ fluorescence scattered by the MOT. While $\sigma^-$ light is used to pump and cycle atoms that are being slowed by our Zeeman slower, the magnetic field magnitude and direction at the position where the angled slowing beams intersect the atomic beam are not easily known, so we varied the polarization of the angled slowing beams to maximize the MOT population after the pointing had been optimized. 

The angled slowing light is prepared by frequency shifting light from our $421\;{\rm nm}$ master laser with a $500\;{\rm MHz}$ AOM in a double-pass configuration, and then splitting the shifted light into two separate fibers. With more than $200\;{\rm mW}$ of input power to this frequency shifting setup, thermal lensing in the AOM causes sensitively power-dependent variations in the spatial mode of the beams reaching the fibers, greatly reducing the fiber coupling efficiency. To avoid thermal lensing (and allow for more controlled variation of the angled slowing power via the RF power), we keep the power going to this AOM low, resulting in a maximum power of about $10\;{\rm mW}$ per beam in our angled slowing light.

Given this power constraint, we looked for an optimal combination of power and detuning for the angled slowing beams. Figure 3 shows the population after a fixed load time and fixed compression sequence (see the following section) as a function of both detuning (always red) and power per beam. The uncertainty in our beam power was at most $\pm 15 \%$, and the uncertainty in our frequency was about $\pm 2\;{\rm MHz}$, with the latter uncertainty arising from drifts in our wavemeter.

The general trends are explained by a simple physical picture: At small detunings, a small amount of power kicks some of the slowed flux to below the capture velocity of the MOT, but increasing power causes significant additional scattering in the nearly-zero-velocity atoms and causes them to turn around. At intermediate detunings, more flux is kicked out of the broad, slowed distribution to velocities below the capture velocity. Eventually, with enough power, off-resonant slowing begins to turn the atoms around again. At large detunings, the majority of the slowed flux is only addressed off-resonantly by the angled slowing beams. Eventually, atoms will also be turned around off-resonantly and thus there should be an optimal power for any given detuning.

At each detuning, we scanned the constant offset field in the slower, which is equivalent to scanning the Zeeman slower laser frequency and hence the final velocity. We found that the optimal bias field was the same for all detunings to within the step size we explored (steps of $1\;{\rm A}$ $\approx \frac{\Gamma_{421}}{2}$ of effective Zeeman slower detuning). We also found that the same bias field was optimal when loading a MOT without angled slowing. Together, these observations suggest that the final velocity distribution of the Zeeman-slowed atoms is broad compared to $\Gamma_{421}$ If the final velocity distribution were narrow, we would expect the optimum to vary with the choice of angled slowing detuning.

\begin{figure}
    \includegraphics[scale=0.4]{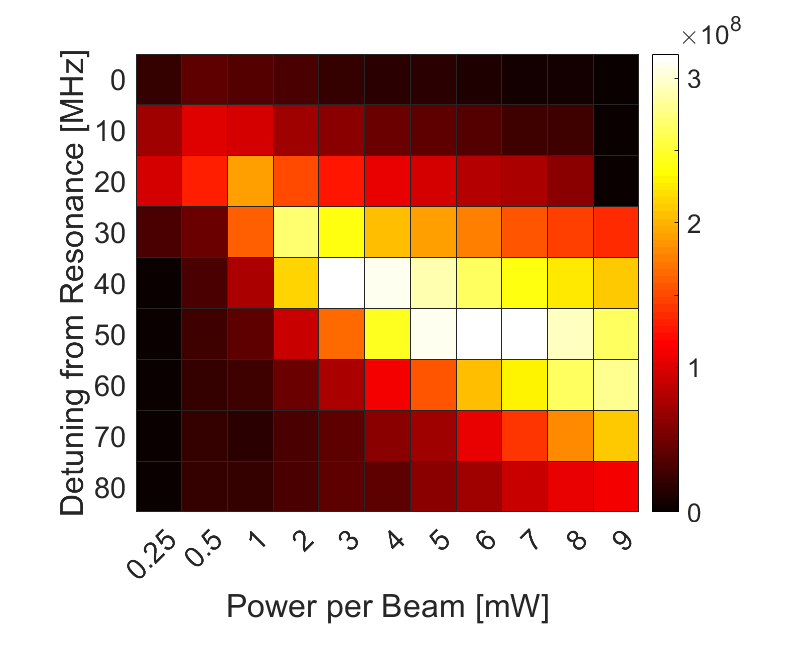}
    \label{fig:optimization}
    \caption{Optimizing angled slowing. Dependence of the MOT population on the detuning and power of the angled slowing beams. A fixed loading time and compression sequence were employed for all of the data shown. The powers in the two beams were balanced to within 15\%. The detuning from the $421\;{\rm nm}$ resonance was known to within $\pm 2\;{\rm MHz}$.}
\end{figure}
\section{Compression and Detection}
We load about $3\times 10^8$ atoms in 2 seconds with our optimized angled slowing parameters. To prepare the captured atoms to be loaded into an optical dipole trap (ODT) for evaporation, we compress the cloud over 50ms and let the compressed cloud equilibrate for at least $300\;{\rm ms}$. Compression consists of switching off the dithering of the MOT light frequency, and ramping the frequency from the initial detuning to within about a few linewidths of resonance. To minimize losses due to light-assisted collisions, and to reach the lowest final temperature of the cloud, the MOT beams are ramped down to a final power of $22\;\mu$W per arm. We also reduce the magnetic field gradient from $2.5\;{\rm G/cm}$ to $1.75\;{\rm G/cm}$ in order to further minimize losses. At the end of the compression, the MOT is approximately $400\;\mu{\rm m} \times 400\;\mu{\rm m} \times 800\;\mu{\rm m}$. We lose up to half of our atoms during the $300\;{\rm ms}$ of post-compression equilibration, but obtain a net gain in phase space density due to the simultaneous reduction in temperature. 

To detect the number of atoms captured in our trap, we perform absorption imaging using light resonant with the $421\;{\rm nm}$ transition. We expect a high degree of spin polarization in the $m_J = -8$ spin state as a result of the force of gravity on our narrow-line MOT as discussed in \cite{ParisMOT}, and so we image using $\sigma^-$ light to address the $m_J = -8 \rightarrow m_J' = -9$ transition, which has a Clebsch-Gordon coefficient of nearly unity \cite{MWLThesis}. We let the cloud expand freely for between 10ms and 30ms before shining a $100\; \mu{\rm s}$ imaging light pulse. We have verified that we have a high degree of spin polarization by using $\sigma^+$ light instead of $\sigma^-$ light, and observing that the optical depth was reduced by more than an order of magnitude.  

We measure the temperature of our cloud after compression by loading successive MOTs with identical parameters and varying the time-of-flight (TOF) after turning off the MOT beams and quadrupole. By fitting the cloud size as a function of the TOF, we can observe the mean speed of the cloud and hence the temperature. We observe faster expansion along the vertical direction than along the horizontal direction, corresponding to a ``vertical temperature'' of $6\;\mu$K and a ``transverse temperature'' of $13\;\mu$K.

To obtain the optimal phase space density, $n \lambda_{T}^3$, we varied the MOT frequency, detuning, and gradient during the compression sequence. We used the size in large TOF ($20-25\;{\rm ms}$) as a proxy for velocity (and therefore temperature), which in combination with the measured number allowed for single-shot estimation of the phase space density. We optimized the phase space density both through manual parameter scans and by automating the search using a genetic algorithm, which converged after about 4 generations. We obtained similar results from both approaches, and measured an optimal phase space density of $10^{-5}$ after 10 seconds of loading. 

\section{Conclusion}
In conclusion, we used the described angled slowing technique to reduce the effect of transverse atomic beam spreading on our MOT loading, effectively increasing the capture velocity of our narrow-line MOT. We observe more than an order of magnitude increase in the number of atoms captured in the MOT when the angled slowing is operated with optimal parameters, allowing us to load MOTs in the $10^8$ regime in a few seconds. In our experiment, the combination of a narrow cooling transition, long free-flight distance, and reduced off-resonant slowing means that the free-flight time is particularly long; we believe that angled slowing can be of use in similarly designed experiments using species with narrow cooling transitions (such as Dy, Er, or Yb). Even in experiments where transverse spread can be avoided by employing other techniques, such as transverse Doppler cooling or the core-shell MOT configuration, the low power requirements and simple geometry of the angled slowing scheme may make it a comparatively attractive option. 

\section{Acknowledgements}
We would like to thank Benjamin L. Lev, Tilman Pfau, and Isaac L. Chuang for fruitful discussions during the early construction stages of our experiment as well as Rudy Pei, Rokas Veitas, and Cody Winkleblack for experimental assistance. We acknowledge support from ONR (Grant No.N00014-17-1-2253), and from a Vannevar-Bush Faculty Fellowship. M.C. acknowledges support by the National Science Foundation Graduate Research Fellowship under Grant No. 1122374.

\providecommand{\noopsort}[1]{}\providecommand{\singleletter}[1]{#1}%

\end{document}